\documentclass[twocolumn,superscriptaddress,showpacs,preprintnumbers,amsmath,amssymb]{revtex4}

\usepackage[dvips]{color}
\usepackage{graphicx}
\usepackage{dcolumn}
\usepackage{bm}

\usepackage{amsmath,amssymb}

\begin{document}

\title{The Higgs mode in a superfluid of Dirac fermions}

\author{Shunji Tsuchiya}
\affiliation{Department of Physics, Faculty of Science, Tokyo University
of Science, 1-3 Kagurazaka, Shinjuku-ku, Tokyo 162-8601, Japan}

\author{R. Ganesh}
\affiliation{Institute for Theoretical Solid State Physics, IFW Dresden,
PF 270116, 01171 Dresden, Germany}

\author{Tetsuro Nikuni}
\affiliation{Department of Physics, Faculty of Science, Tokyo University
of Science, 1-3 Kagurazaka, Shinjuku-ku, Tokyo 162-8601, Japan}

\date{\today}

\begin{abstract}
We study the Higgs amplitude mode in the $s$-wave superfluid state on the honeycomb lattice inspired
by recent cold atom experiments. We consider the attractive Hubbard model and focus on the vicinity of a quantum phase transition between semi-metal and superfluid phases.
On either side of the transition, we find collective mode excitations that are stable against decay into quasiparticle-pairs. In the semi-metal phase, the collective modes have ``Cooperon'' and exciton character. These modes smoothly evolve across the quantum phase transition, and become the Anderson-Bogoliubov mode and the Higgs mode of the superfluid phase.
The collective modes are accommodated within a window in the quasiparticle-pair continuum, which arises as a consequence of the linear Dirac dispersion on the honeycomb lattice, and allows for sharp collective excitations.
Bragg scattering can be used to measure these excitations in cold atom experiments, providing a rare example wherein collective modes can be tracked across a quantum phase transition.
\end{abstract}

\maketitle
\textit{Introduction}--
Spontaneous symmetry breaking of continuous symmetries gives rise to two
typical collective excitations - gapless Goldstone modes and a gapped
amplitude mode, also called the Higgs mode~\cite{Varma}.
While the Goldstone mode has been observed in various contexts, the Higgs mode has evaded observation with rare exceptions such as
NbSe$_2$, which has coexisting charge density wave and
superconducting order~\cite{Sooryakumar,Littlewood} and multiferroic Ba$_2$CoGe$_2$O$_7$~\cite{Penc}.
Remarkably, two recent experiments have successfully observed this mode by tracking collective excitations across a quantum phase transition.
The first involves pressure studies of TlCuCl$_3$, a magnetic material
which undergoes a transition from dimer order to magnetic
order~\cite{Ruegg}.
The second is the realization of the Bose-Hubbard model in ultracold
gases, with a visible amplitude mode near the superfluid-Mott
transition~\cite{Bissbort,Endres}. In this letter, we propose a novel scheme to
observe the Higgs mode in a \textit{Fermi} superfluid. Hitherto, the
Higgs mode has never been observed in Fermi superfluids as it
decays into pairs of quasiparticles. Our proposal circumvents this issue
by exploiting a special feature of the honeycomb lattice geometry which
allows for a window in the quasiparticle-pair continuum -- the Higgs
mode survives as a stable excitation inside this window.
\par
Inspired by the recent realization of the honeycomb optical lattices in cold atom experiments~\cite{Tarruell}, we study the attractive Hubbard model in this geometry:
\begin{equation}
H=-\sum_{i,j,\sigma}t_{ij}c_{i\sigma}^\dagger
c_{j\sigma}-\mu\sum_{i,\sigma}n_{i\sigma}-U\sum_i
n_{i\uparrow}n_{i\downarrow}~.
\label{eq.Hubbard}
\end{equation}
Parameter $t_{ij}$ denotes hopping amplitude between nearest neighbor
($t_{ij}=t$) and next-nearest neighbor sites ($t_{ij}=t'$). $U$ is an on-site attractive interaction and $\mu$ is the chemical
potential.
We envisage a setup with a deep optical lattice to trap two hyperfine
species of fermions, and a magnetic field on the attractive side of a
Feshbach resonance~\cite{Bloch}.
This model hosts a superfluid state of Dirac fermions, with several
interesting implications~\cite{Zhao,Tsuchiya}.
In this proposal, we make use of two key features: (i) strictly at
half-filling, there is an interaction-tuned quantum phase transition
from a semi-metal phase to an $s$-wave superfluid. This has been
demonstrated by sophisticated quantum Monte Carlo simulations on very
large system sizes~\cite{Sorella,p-hmapping}.
This transition is a consequence of the Dirac cone dispersion which leads to vanishing density of states at the Fermi level, thereby necessitating a critical interaction strength to induce superfluid order~\cite{Nozieres, Zhao}.
(ii) In the semi-metal phase, the two-particle continuum has a window
structure, again a consequence of the Dirac cone dispersion~\cite{Baskaran}. A collective mode excitation propagating inside
this window is stable against decay into quasiparticle-pairs. We show
that this window structure persists in the superfluid phase, thus
allowing for a stable Higgs mode excitation.
\par
\begin{figure}
\centerline{\includegraphics[width=\linewidth]{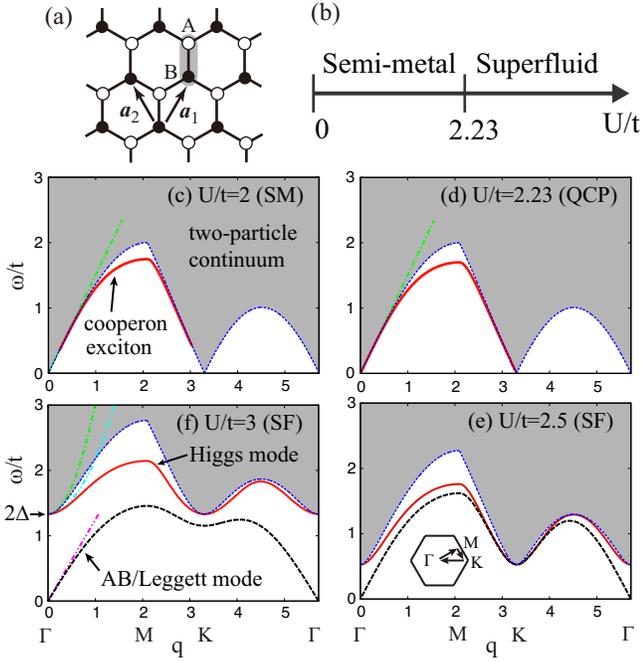}}
\caption{Isotropic honeycomb lattice with basis vectors (a). Phase
 diagram of the attractive Hubbard model on a honeycomb
 lattice at half-filling with $t'=0$ as obtained from the mean-field theory (b).
 Evolution of the elementary excitations across the quantum critical
 point (QCP) of the semi-metal (SM) to superfluid (SF) phase transition (c)-(f).
 The dash-dot line in each panel and the dash-dot-dot lines in (f) show the asymptotic
 dispersions of the continuum edge and the superfluid collective modes
 for small $q$, respectively.
 }
\label{fig1}
\end{figure}
Fig.~\ref{fig1}(b) shows the phase diagram of this model at half-filling. Our key
findings are summarized in Fig.~\ref{fig1}(c)-(f): (i) there are sharp collective mode
excitations on either side of the transition. The two-particle
continuum is shown as the shaded region, note the window
structure.
(ii) In the semi-metal phase, there are three degenerate collective modes with ``Cooperon'' and exciton character.
(iii) On the superfluid side, there is a Goldstone mode and remarkably, a distinct superfluid amplitude (Higgs) mode. These modes can be observed in a cold
atoms experiment using Bragg scattering. This is a rare
example wherein relevant collective excitations can be tracked across a
quantum phase transition.
\par
\textit{Mean field theory}--
The ETH group~\cite{Tarruell} has studied fermions loaded onto a honeycomb optical lattice with tunable anisotropy.
We consider the attractive Hubbard
model in the \textit{isotropic} honeycomb lattice. As discussed in
Ref.~\cite{Tsuchiya}, the isotropic limit is expected
to have the highest superfluid transition temperature and is the most promising
for experimental realization. We decompose the Hubbard interaction in
the superfluid channel using the order parameter $\Delta=U\langle
c_{i\downarrow}c_{i\uparrow}\rangle$, taken to be real.
For brevity, we introduce a vector operator consisting of
creation and annihilation operators $\hat\Psi(\bm p)=(
c_{\bm p,a,\uparrow},
c^\dagger_{-\bm p,a,\downarrow},
c_{\bm p,b,\uparrow},
c^\dagger_{-\bm p,b,\downarrow})^t$ ($a$ and $b$ denote the two sublattices as
shown in Fig.~\ref{fig1}(a)). The mean field Hamiltonian can be
written as $H_{\rm MF}=\sum_{\bm p}\hat\Psi^\dagger(\bm p)\hat h(\bm p)\hat\Psi(\bm
p)$,
where $\hat h(\bm p)=\{x_{\bm p}+{\rm Re}(\gamma_{\bm
p})\sigma_x-{\rm Im}(\gamma_{\bm p})\sigma_y\}\tau_3-\Delta\tau_1$,
$\gamma_{\bm p}=-t(1+e^{i\bm p\cdot \bm a_1}+e^{i\bm p\cdot\bm a_2})$ and $x_{\bm p}=-2t'[\cos(\bm p\cdot\bm a_1)+\cos(\bm p\cdot\bm a_2)+\cos(\bm p\cdot
(\bm a_1-\bm a_2))]-\mu$ with $\bm a_1$ and $\bm a_2$ being the two basis
vectors shown in Fig.~\ref{fig1} (a). We take the lattice spacing to be
unity. $\vec \tau$ and $\vec \sigma$ are the
Pauli matrices in the Nambu and sublattice space, respectively.
The single-particle Green's function for the mean-field Hamiltonian is given by
\begin{eqnarray}
\hat G(p)=\left[i\omega_n-\hat h(\bm
	     p)\right]^{-1}
\equiv\left(
\begin{array}{cc}
\hat G^{aa}(p) & \hat G^{ab}(p)\\
\hat G^{ba}(p) & \hat G^{bb}(p)
\end{array}
\right).\label{greenf}
\end{eqnarray}
Here, we denote $p=(\bm p,i\omega_n)$, where $\omega_n$
is the fermion Matsubara frequency.
The gap and number equations are obtained from the off-diagonal and
diagonal elements of the Green's function ${\hat G}^{\nu\nu}$ as~\cite{Zhao}
(hereafter, we restrict ourselves to zero temperature)
\begin{eqnarray}
\frac{1}{U}&=&\frac{1}{N}\sum_{\bm p}\sum_{\alpha=\pm}\frac{1}{2E^\alpha(\bm p)},\label{gapeq}\\
n&=&1-\frac{1}{N}\sum_{\bm p}\sum_{\alpha=\pm}\frac{\xi_{\bm p}^\alpha}{E^\alpha(\bm p)},\label{numbereq}
\end{eqnarray}
where $E^\pm(\bm p)=\sqrt{(\xi_{\bm p}^\pm)^2+\Delta^2}$ is the spectrum
of the Bogoliubov quasiparticles, $\xi_{\bm p}^\pm=x_{\bm
p}\pm|\gamma_{\bm p}|$, and $N$ is the number of lattice sites.
At half-filling, the self-consistent solution of $\Delta$ becomes
non-zero for $U> U_c$ indicating a transition from semi-metal to
superfluid phases~\cite{Zhao,Tsuchiya}. For $t'=0$, mean-field theory gives $U_c \sim
2.23t$. Quantum Monte Carlo gives the same transition, except with $U_c$
renormalized to $\sim 3.869$~\cite{Sorella}. In the rest of this letter, we
use mean-field results with the understanding that fluctuations will
renormalize $U$ quantitatively.
We note that $U_c$ weakly depends on the value of $t'$.
\par
\textit{Generalized Random Phase Approximation (GRPA)}--
On either side of the critical point, there are low-lying density and pairing
fluctuations. We use a generalized random phase approximation (GRPA) scheme
to evaluate density and pairing response functions. We follow the
Green's function approach of C\^ot\'e and Griffin~\cite{Cote}.
We denote matrix susceptibilities containing the response to weak
density and pairing perturbations, respectively, as
\begin{eqnarray}
\hat L^{\nu_1\nu_2}(q)=
\left(
\begin{array}{cc}
\chi^{\nu_1\nu_2}_{n_{\uparrow}n}(q) & \chi^{\nu_1\nu_2}_{mn}(q) \\
-\chi^{\nu_1\nu_2}_{m^\dagger n}(q) & \chi^{\nu_1\nu_2}_{n_{\downarrow} n}(q)
\end{array}
\right),\\
\hat M^{\nu_1\nu_2}(q)=
\left(
\begin{array}{cc}
\chi^{\nu_1\nu_2}_{n_{\uparrow}m^\dagger}(q) & \chi^{\nu_1\nu_2}_{mm^\dagger}(q) \\
-\chi^{\nu_1\nu_2}_{m^\dagger m^\dagger}(q) & \chi^{\nu_1\nu_2}_{n_{\downarrow}m^\dagger}(q)
\end{array}
\right),
\end{eqnarray}
where $q=(\bm q,i\Omega_n)$ ($\Omega_n$ is a boson Matsubara frequency).
Any susceptibility $\chi$ is defined as
\begin{equation}
\chi_{fg}^{\nu_1\nu_2}(q)=-\sum_{\bm r_{12}} \int_0^\beta
 d\tau_{12}\langle T_\tau\delta f(1)\delta g(2)\rangle e^{-i(\bm q\cdot\bm r_{12}-\Omega_n\tau_{12})},
\end{equation}
where $1\equiv(\bm r_{l_1},\nu_1,\tau_1)$ ($l_1$ denotes the unit cell,
$\nu_1$ the sublattice, and $\tau_1$ an imaginary time), $\bm r_{12}=\bm r_{l_1}-\bm r_{l_2}$,
$\tau_{12}=\tau_1-\tau_2$ and $\delta f\equiv f-\langle
f\rangle$. The density and pair annihilation operators are written as
$n=n_{\uparrow}+n_{\downarrow}$ and
$m=c_{\downarrow}c_{\uparrow}$, respectively.
The GRPA equations read~\cite{Tsuchiya,Cote}
\begin{eqnarray}
{\bar A}^{\nu_1\nu_2}(q)&=&\hat A^{0\nu_1\nu_2}(q)+\frac{2U}{\beta
 N}\sum_{\nu_3}\sum_{\bm p,\omega_n}{\tilde G}^{\nu_1\nu_3}(p+q)\nonumber\\
&&\times{\bar A}^{\nu_3\nu_2}(q){\tilde
 G}^{\nu_3\nu_1}(p),\label{GRPA2}\\
\hat{A}^{\nu_1\nu_2}(q)\!&=& \!\!\!{\bar A}^{\nu_1\nu_2}(q)\!-\! U\!\sum_{\nu_3}{\bar L}^{\nu_1\nu_3}(q) \mathrm{Tr}\{ \hat{A}^{\nu_3\nu_2}(q)\},\label{GRPA1}
\end{eqnarray}
where $A$ is either $L$ or $M$ and ${\tilde G}^{\nu_1\nu_2}(p)=\tau_3
\hat G^{\nu_1\nu_2}(p)$. $A^0$ denotes the bare susceptibility~\cite{bareA0}, $\bar{A}$ includes an infinite sum over ladder diagrams, while $\hat{A}$ is the final result which also includes bubble diagrams.
\par
\textit{Undamped Higgs mode}--
In the superfluid phase, we solve GRPA equations (\ref{GRPA2}) and (\ref{GRPA1}) to
evaluate the amplitude and phase correlation functions $\chi_{\Delta\Delta}^{\nu_1\nu_2}(q)
=\frac{U^2}{2}(\chi_{mm^\dagger}^{\nu_1\nu_2}(q)+\chi_{m^\dagger
m^\dagger}^{\nu_1\nu_2}(q))$ and
$\chi_{\theta\theta}^{\nu_1\nu_2}(q)=\frac{U^2}{2\Delta^2}(\chi_{mm^\dagger}^{\nu_1\nu_2}(q)-\chi_{m^\dagger
m^\dagger}^{\nu_1\nu_2}(q))$. The amplitude and phase fluctuation operators are given by
$\delta\Delta=\frac{U}{2}(\delta m+\delta m^\dagger)$ and
$\delta\theta=\frac{U}{2i\Delta}(\delta m-\delta m^\dagger)$,
respectively. For the case of $t'=0$, the expressions simplify
and we can identify their respective poles, which we denote ``Higgs'' and ``AB/Leggett''. These poles satisfy
\begin{eqnarray}
Higgs:\phantom{a}\frac{1}{U}&=&-(C+D)+|R|,\label{poleHiggs}\\
AB/Leggett:\phantom{a}\frac{1}{U}&=&-(C-D)+\sqrt{4F^2+|R|^2}.\ \  
\label{poleAB}
\end{eqnarray}
We have defined
\begin{eqnarray}
C&=&\frac{1}{N}\sum_{\bm p}\frac{E+E'}{(i\Omega_n)^2-(E+E')^2},\\
D&=&-\frac{1}{N}\sum_{\bm
p}\frac{\Delta^2}{E'E}\frac{E+E'}{(i\Omega_n)^2-(E+E')^2},\\
F&=&\frac{1}{N}\sum_{\bm
p}\frac{\Delta}{E}\frac{i\Omega_n}{(i\Omega_n)^2-(E+E')^2},\\
R&=&-\frac{1}{N}\sum_{\bm p}\frac{\gamma'\gamma^*}{E'E}\frac{E+E'}{(i\Omega_n)^2-(E+E')^2}.
\end{eqnarray}
Here, we have denoted $E=E(\bm p)$, $E'=E(\bm p+\bm q)$, $\gamma=\gamma_{\bm
p}$, and $\gamma'=\gamma_{\bm p+\bm q}$.
On the other hand, solving Eqs.~(\ref{GRPA2}) and (\ref{GRPA1}) for
density response, we find that
$\chi_{\theta\theta}^{\nu_1\nu_2}(q)\propto\chi_{nn}^{\nu_1\nu_2}(q)$ when $t'=0$. Thus, the density response function only retains the AB/Leggett pole
given in Eq.~(\ref{poleAB}).
\par
Setting $\bm q=\Omega_n=0$ in Eq.~(\ref{poleAB}), we recover the gap
equation (\ref{gapeq}). Thus, the superfluid phase has {\it gapless} collective
mode(s) arising from phase fluctuations.
In fact, at half-filling, the AB/Leggett pole in
Eq.~(\ref{poleAB}) is a double pole corresponding to two gapless modes: the
Anderson-Bogoliubov (AB) mode and the Leggett mode~\cite{Tsuchiya}.
The AB mode is the usual Goldstone mode associated with U(1)
symmetry breaking~\cite{Anderson,Nambu}.
The Leggett mode is composed of out-of-phase fluctuations between sublattices~\cite{Leggett} - it acquires a gap away from half-filling~\cite{Tsuchiya}.
The AB and Leggett modes become degenerate at half-filling reflecting a special pseudospin SU(2) symmetry of the Hubbard model~\cite{Tsuchiya}.
For small $q\ll 1$, Eq.~(\ref{poleAB}) gives the dispersion relation of
the AB/Leggett mode to be~\cite{Supple}
\begin{eqnarray}
\omega_{\rm AB}=\lambda v_Fq,\ \
\lambda^2=\frac{U}{N}\sum_{\bm p}\frac{|\gamma|^2}{E^3}\leq 1,
\label{ABdispersion}
\end{eqnarray}
where $v_F=3t/2$ is the Fermi velocity at the Dirac points.
\par
Setting $(\bm q,i\Omega_n)=(0,2\Delta)$, Eq.~(\ref{poleHiggs}) also
reduces to the gap equation. Thus, there exists a {\it gapped} collective
mode with the energy gap $2\Delta$ at $\bm q=0$.
This is the `Higgs' mode or the amplitude mode~\cite{Littlewood} arising from
amplitude fluctuations of the superfluid order parameter. It can be understood using the mechanical analog
of motion along the radial direction of the famous ``Mexican hat''
potential; the energy gap stems from the finite curvature of the potential
along the radial direction.
\par
Remarkably, the Higgs mode disperses below the
quasiparticle pair continuum in Figs.~\ref{fig1}(e) and (f). In particular, close to the $M$ point, it is well separated from the lower edge of the continuum.
This is to be contrasted with the case of typical superfluids:
due to the underlying Fermi surface, the continuum exhibits a horizontal edge
near $q\sim 0$~\cite{Littlewood}. The Higgs mode therefore enters the continuum, becomes heavily damped and is unobservable.
\par
In our case, the Higgs mode in Fig.~\ref{fig1} is undamped over large sections of the Brillouin zone.
For $q\ll 1$, solving Eq.~(\ref{poleHiggs}), the Higgs mode has the dispersion relation $\omega_{\rm Higgs}^2=4\Delta^2+v_F^2q^2$~\cite{Supple}.
The Higgs mode disperses below the lower edge of the continuum which is given by
$\omega_{\rm edge}^2={\rm min}_{\bm
p}[(E(\bm p)+E(\bm p+\bm q))^2]\simeq 4\Delta^2+2v_F^2q^2$ ($q\ll 1$).
This window or arch in the continuum, shown in Fig.~\ref{fig1}, is a consequence of the Dirac-like dispersion of underlying fermions. The Higgs mode stays undamped as long as it lies within this window.
Even if we go slightly away from half-filling, the Higgs mode survives undamped. Close to $n\sim 0.9$ or $n\sim 1.1$, the window disappears because of the
presence of the Fermi surface - as a result, the Higgs mode is strongly
damped.
The Higgs mode, the AB mode, and the lower edge of the continuum become
degenerate at the QCP for $q\ll 1$: $\omega_{\rm AB}=\omega_{\rm
Higgs}=\omega_{\rm edge}=v_Fq$.
\par
\begin{figure}
\centerline{\includegraphics[width=\linewidth]{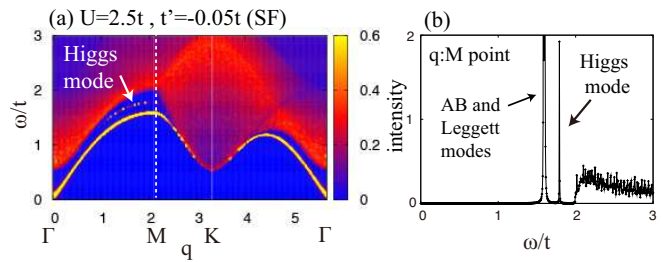}}
\caption{Intensity of dynamic structure factor corresponding to density
 response $S(\bm q,\omega)=-{\rm
 Im}[\chi_{nn}(\bm q,\omega)]/\pi$ for $t'=-0.05t$ (a).
 The cross section for the momentum at the $M$ point (b). The upper peak
 corresponds to the Higgs mode.}
\label{fig2}
\end{figure}

The AB mode and Leggett mode are strongly coupled with density
fluctuations; they appear as poles in the density response function
$\chi_{nn}^{\nu_1\nu_2}(q)$.
However, when $t'=0$, the Higgs mode has no corresponding pole in the density
response function. Thus, the Higgs mode is composed of pure amplitude
fluctuations and cannot be excited by
a density perturbation.
This reflects the underlying SU(2) pseudospin symmetry~\cite{Tsuchiya} in the problem.
A small finite $t'$ breaks this symmetry and forces the Higgs
mode to acquire a density component: the density response then shows a peak at the Higgs mode, as shown in Fig.\ref{fig2}.
\par
\textit{Collective modes in the semi-metal phase: Cooperons and excitons}--
In the semi-metal phase, setting $\Delta=0$, density and pair response
functions become decoupled in the GRPA equations (\ref{GRPA2})-(\ref{GRPA1}).
The susceptibilities satisfy the usual RPA equations
$\chi_{mm^\dagger}^{\nu_1\nu_2}(q)=\chi_{mm^\dagger}^{0\nu_1\nu_2}(q)+U\sum_{\nu_3}\chi_{mm^\dagger}^{0\nu_1\nu_3}(q)\chi_{mm^\dagger}^{\nu_3\nu_2}(q)$,
$\chi_{n_\sigma n}^{\nu_1\nu_2}(q)=\chi_{n_\sigma
n}^{0\nu_1\nu_2}(q)-U\sum_{\nu_3}\chi_{n_\sigma
n}^{0\nu_1\nu_3}(q)\chi_{n_{-\sigma}n}^{\nu_3\nu_2}(q)$.
The bare susceptibility $\chi_{mm^\dagger}^0$ describes a single rung diagram
with particle-particle (hole-hole) excitations, and $\chi_{n_\sigma
n}^0$ describes a single bubble diagram with particle-hole excitations.
They are given by
\begin{eqnarray}
&\chi_{mm^\dagger}^{0\nu_1\nu_2}(q)=\frac{1}{2N}\sum_{\bm
 p}\left[\frac{\kappa^{\nu_1\nu_2}_{\bm p}\kappa^{\nu_1\nu_2}_{\bm q-\bm p}}{\xi_{\bm p}^+
    +\xi_{\bm q-\bm p}^+-i\Omega_n}-\frac{\eta^{\nu_1\nu_2}_{\bm
    p}\eta^{\nu_1\nu_2}_{\bm q-\bm p}}{\xi_{\bm p}^-+\xi_{\bm q-\bm
    p}^--i\Omega_n}\right],\ \ \ \label{bareM}\\
&\chi_{n_\sigma n}^{0\nu_1\nu_2}(q)=\frac{1}{2N}\sum_{\bm p}\left[\frac{-\kappa^{\nu_1\nu_2}_{\bm
    p+\bm q}\eta^{\nu_2\nu_1}_{\bm p}}{\xi_{\bm p+\bm q}^+ -\xi_{\bm
    p}^--i\Omega_n}+\frac{\eta_{\bm p+\bm q}^{\nu_1\nu_2}\kappa_{\bm
    p}^{\nu_2\nu_1}}{\xi_{\bm p+\bm q}^- -\xi_{\bm
    p}^+-i\Omega_n}\right],\ \ \ \label{bareL}
\end{eqnarray}
where $\kappa_{\bm p}^{\nu_1\nu_2}=\delta_{\nu_1\nu_2}+e^{i\phi_{\bm
p}}\delta_{\nu_1a}\delta_{\nu_2b}+e^{-i\phi_{\bm
p}}\delta_{\nu_1b}\delta_{\nu_2a}$ and $\eta_{\bm
p}^{\nu_1\nu_2}=\delta_{\nu_1\nu_2}-e^{i\phi_{\bm
p}}\delta_{\nu_1a}\delta_{\nu_2b}-e^{-i\phi_{\bm
p}}\delta_{\nu_1b}\delta_{\nu_2a}$. We denote $e^{i\phi_{\bm
p}}=\gamma_{\bm p}/|\gamma_{\bm p}|$.
\par
From the denominator in Eq.~(\ref{bareM}),
we see that pairing response arises from particle--particle (or
hole--hole) excitations. An undamped pairing mode, occurring below the
particle--particle continuum in Fig.~\ref{fig1}(c), is therefore a two
particle bound state with well defined momentum and energy. It
can be understood as a preformed Cooper pair -- we call this a
\textit{Cooperon} excitation~\cite{Rice}.
Similarly, from Eq.~(\ref{bareL}), we see that density response arises
from particle--hole excitations. An undamped density mode is thus a particle-hole bound state -- we call this an \textit{exciton}.
\par
The dispersions of Cooperons and excitons are determined by the
poles of the corresponding response functions, giving
$|\hat I-U\hat \chi_{mm^\dagger}(q)|=0$ and $|\hat I+U\hat \chi_{n_{\sigma}n}^0(q)|=0$.
With $t'=0$, these reduce to the identical equation
\begin{eqnarray}
&&(1+U\alpha(q))^2-U^2|\beta(q)|^2=0,\label{pole_cooperon}\\
&&\alpha(q)=\frac{1}{N}\sum_{\bm p}\frac{|\gamma_{\bm p}|+|\gamma_{\bm
 p+\bm q}|}
{(i\Omega_n)^2-(|\gamma_{\bm p}|+|\gamma_{\bm p+\bm q}|)},\label{alpha}\\
&&\beta(q)=-\frac{1}{N}\sum_{\bm p}\frac{e^{i(\phi_{\bm p+\bm
 q}-\phi_{\bm p})}(|\gamma_{\bm p}|+|\gamma_{\bm p+\bm
 q}|)}{(i\Omega_n)^2-(|\gamma_{\bm p}|+|\gamma_{\bm p+\bm q}|)^2}.\label{beta}
\end{eqnarray}
Thus, the Cooperon and the exciton are degenerate when $t'=0$.
Their dispersion is shown in Figs.~\ref{fig1} (c) and (d) -- the
modes are undamped as they lie below the two particle continuum.
In particular, they are well separated from the continuum in the vicinity of the $M$ points. We suggest that experiments should probe this region to observe the collective excitations. This feature of the $M$ points can be understood from the single particle band structure which has saddle points at these wavevectors. They consequently have a very large density of
states which provides large phase space for the Hubbard interaction to
form two--particle bound states.
\par
These collective modes in the semi-metal phase were predicted many years ago -- using an insightful single-cone approximation -- Ref.~\cite{Baskaran} reported a triplet exciton mode in the repulsive Hubbard model. The authors identified the window structure in the continuum as capable of accommodating stable modes. Mapping their results to the attractive Hubbard case~\cite{p-hmapping}, the triplet excitons translate to Cooperon and exciton modes.
We reaffirm their prediction starting from a microscopic picture taking into account the sublattice structure.
Our expressions also agree with those of Ref.~\cite{Peres} -- which only considers the
$\Gamma-K$ segment and concludes that there is no undamped
mode. However, we find an undamped mode in the $\Gamma-M$ and $M-K$ directions.
\par
\textit{Cooperon condensation}--
As we approach the critical point from the semi-metal side, the energy
of the Cooperon and exciton decreases progressively (see
Fig.~\ref{fig1}(c) and (d)). Precisely at the transition, the Cooperon
``softens'' at $\bm q=0$ and undergoes condensation.
In fact, setting $\bm q=\Omega_n=0$, the Cooperon pole in Eq.~(\ref{pole_cooperon}) reduces to the
gap equation~(\ref{gapeq}).
Since Cooperons and excitons are degenerate for $t'=0$, the exciton can
also condense at the critical point. That gives rise to the
sublattice--CDW state - which is degenerate with the superfluid state due to SU(2) pseudospin symmetry. For
$t'\neq 0$, this degeneracy is lifted in favour of the superfluid and the Cooperon condenses preferentially.
\par
As we cross $U_c$ and enter the superfluid phase, Cooperons and excitons
hybridize to become the AB, Leggett and Higgs modes. The excitonic
component, when present, allows these modes to have peaks in
the density response function. 
The Cooperonic component manifests as peaks in the pairing response.
Thus, the collective modes evolve smoothly across the QCP and carry
signatures of the underlying spontaneous symmetry breaking.
\par
\textit{Visibility of the Higgs mode}--
Our Higgs mode is stable against decay into pairs of fermions due to the
window structure in the two particle continuum. However,
if we go beyond RPA level, the Higgs mode may decay
by emitting AB/Leggett excitations which are lower in energy. We argue that
this merely leads to broadening of the Higgs mode. In our semi-metal to
superfluid transition, due to the pseudospin symmetry present when $t'=0$, the order parameter can be thought of as an O(3) object.
Quantum Monte Carlo simulations of the O(3) ordering transition show
that the Higgs mode survives although it is broadened by decay into
Goldstone bosons~\cite{Gazit}. We expect this to be true in our case as well.
\par
We suggest Bragg spectroscopy measurements on a Fermi superfluid in a honeycomb
optical lattice as a way to measure this undamped Higgs
mode. In this technique, a two-photon process imparts a density-``kick'' to the system.
The response to this perturbation can be quantified by measuring the momentum transferred or the energy absorbed. The momentum transferred is a measure of the dynamic structure factor
related to the density response function $S(\bm q,\omega)=-{\rm
Im}[\chi_{nn}(\bm q,\omega)]/\pi$~\cite{Pitaevskii} -- it can detect collective modes as long as they have a density component.
For any small $t'\neq 0$, the Higgs mode in our model has a
density component which makes it visible to Bragg spectroscopy.
A small $t'$ hopping is expected to be present in an optical lattice
setup any way~\cite{Ibanez}.
Figure~\ref{fig2} shows $S(\bm q,\omega)$ for $t'=-0.05$. The
sharp intensity peak for the Higgs mode can be clearly seen below the
continuum.
An alternative approach is to measure the energy absorption in response
to a weak shaking of the optical lattice~\cite{Endres}.
\par
\textit{Acknowledgements}--We acknowledge A. Paramekanti for fruitful
dicussions. S.T. thanks M. Sigrist, T. Esslinger, L. Tarruell,
Y. Ohashi, S. Okada, S. Konabe, K. Asano, S. Kurihara, and K. Kamide for discussions.
S. T. was supported by Grant-in-Aid for Scientific Research, No. 24740276.

\clearpage

\section{Supplementary Materials}
\subsection{Bare susceptibility in GRPA}
To solve the GRPA equations (8) and (9), the bare susceptibilities $L^0$
and $M^0$ are evaluated using the mean-field Green's 
function in Eq.~(2) to give
\begin{eqnarray}
&&\hat L^{0\nu_1\nu_2}(q)=\frac{1}{\beta M}\sum_{\bm
p,\omega_n}{\tilde G}^{\nu_1\nu_2}(p+q) {\tilde G}^{\nu_2\nu_1}(p)\nonumber\\
&&=
\left(
\begin{array}{cc}
L^{0\nu_1\nu_2}_{1111}+L^{0\nu_1\nu_2}_{1221} &
 L^{0\nu_1\nu_2}_{1112}+L^{0\nu_1\nu_2}_{1222} \\
L^{0\nu_1\nu_2}_{2111}+L^{0\nu_1\nu_2}_{2221} &
 L^{0\nu_1\nu_2}_{2112}+L^{0\nu_1\nu_2}_{2222}
\end{array}
\right),\\
&&\hat M^{0\nu_1\nu_2}(q)=\frac{1}{\beta M}\sum_{\bm
p,\omega_n}{\tilde G}^{\nu_1\nu_2}(p+q)
\left(
\begin{array}{cc}
0 & 1 \\
0 & 0
\end{array}
\right)
{\tilde G}^{\nu_2\nu_1}(p)\nonumber\\
&&=
\left(
\begin{array}{cc}
L^{0\nu_1\nu_2}_{1121} & L^{0\nu_1\nu_2}_{1122} \\
L^{0\nu_1\nu_2}_{2121} & L^{0\nu_1\nu_2}_{2122}
\end{array}
\right),
\end{eqnarray}
where we introduced the tensor 
\begin{equation}
L^{0\nu_1\nu_2}_{ijkl}(q)=\frac{2}{\beta N}\sum_{\bm
 p,\omega_n}{\tilde G}_{ij}^{\nu_1\nu_2}(p+q){\tilde G}_{kl}^{\nu_2\nu_1}(p).
\end{equation}
Following Ref.~\cite{Cote}, we introduce a column vector $\mathcal A$($={\mathcal L}, {\mathcal M}$)
and a $4\times 4$ matrix $\mathcal D$ as
\begin{eqnarray}
&&\mathcal{A}=
\left(
\begin{array}{c}
A_{11} \\
A_{12} \\
A_{21} \\
A_{22}
\end{array}
\right)\equiv
\left(
\begin{array}{c}
A_1 \\
A_2 \\
A_3 \\
A_4
\end{array}
\right),\\ 
&&\hat{\mathcal{D}}=
\left(
\begin{array}{cccc}
L^0_{1111} & L^0_{1121} & L^0_{1211} & L^0_{1221}\\
L^0_{1112} & L^0_{1122} & L^0_{1212} & L^0_{1222}\\
L^0_{2111} & L^0_{2121} & L^0_{2211} & L^0_{2221}\\
L^0_{2112} & L^0_{2122} & L^0_{2212} & L^0_{2222}
\end{array}
\right).
\end{eqnarray}
The GRPA equations are cast into the form
\begin{eqnarray}
&&\bar{\mathcal A}^{\nu_1\nu_2}(q)={\mathcal A}^{0\nu_1\nu_2}(q)+U\sum_{\nu_3}\hat{\mathcal D}^{\nu_1\nu_3}(q)\bar{\mathcal A}^{\nu_3\nu_2}(q),\\
&&\mathcal A^{\nu_1\nu_2}(q)=\bar{\mathcal
 A}^{\nu_1\nu_2}(q)-U\sum_{\nu_3}\bar{\mathcal
 L}^{\nu_1\nu_3}(q)A^{\nu_3\nu_2}(q).
\end{eqnarray}
The above equations are easily solved to give
\begin{widetext}
\begin{eqnarray}
\left(
\begin{array}{cc}
A^{aa}(q) & A^{ab}(q) \\
A^{ba}(q) & A^{bb}(q)
\end{array}
\right)
=
\left(
\begin{array}{cc}
1+U{\bar L}^{aa}(q) & U{\bar L}^{ab}(q) \\
 U{\bar L}^{ba}(q) & 1+U{\bar L}^{bb}(q)
\end{array}
\right)^{-1}
\left(
\begin{array}{cc}
\bar A^{aa}(q) & \bar A^{ab}(q) \\
\bar A^{ba}(q) & \bar A^{bb}(q)
\end{array}
\right),\\
\left(
\begin{array}{cc}
\bar{\mathcal A}^{aa}(q) & \bar{\mathcal A}^{ab}(q) \\
\bar{\mathcal A}^{ba}(q) & \bar{\mathcal A}^{bb}(q)
\end{array}
\right)
=
\left(
\begin{array}{cc}
\hat I-U\hat{\mathcal D}^{aa}(q) & -U\hat{\mathcal D}^{ab}(q) \\
-U\hat{\mathcal D}^{ba}(q) & \hat I-U\hat{\mathcal D}^{bb}(q)
\end{array}
\right)^{-1}
\left(
\begin{array}{cc}
{\mathcal A}^{0aa}(q) & {\mathcal A}^{0ab}(q) \\
{\mathcal A}^{0ba}(q) & {\mathcal A}^{0bb}(q)
\end{array}
\right).
\end{eqnarray}
\end{widetext}
Here, we denoted $A^{\nu_1\nu_2}={\rm Tr}\{\hat A^{\nu_1\nu_2}\}={\mathcal A}_1^{\nu_1\nu_2}+{\mathcal A}_4^{\nu_1\nu_2}$. 

\subsection{Energy dispersion of the Higgs mode}

We derive the analytic expression of the energy dispersion of the Higgs
mode for small momentum $ q\ll 1$ following the approach of Ref.~\cite{Littlewood}.
The gap equation~(\ref{gapeq}) can be rewritten as
\begin{equation}
1-\frac{U}{2N}\sum_{\bm p}\left(\frac{1}{E}+\frac{1}{E'} \right)=0.
\label{gapeq2}
\end{equation}
We denote $E(\bm p)=E$, $E(\bm p+\bm q)=E'$, $\gamma=\gamma_{\bm p}$,
and $\gamma'=\gamma_{\bm p+\bm q}$. 
Subtracting Eq.~(\ref{gapeq2}) from $1+U[(C+D)-|R|]=0$ which is
equivalent to Eq.~(\ref{poleHiggs}), we obtain 
\begin{equation}
U\left(X-|R|\right)=0,
\label{Higgs_q}
\end{equation}
where 
\begin{eqnarray}
&&X=(C+D)+\frac{1}{2N}\sum_{\bm p}\frac{E+E'}{EE'}\nonumber\\
&&=\frac{1}{2N}\sum_{\bm
 p}\frac{E+E'}{EE'}\frac{\omega^2-(|\gamma|^2+|\gamma'|^2)-4\Delta^2}{\omega^2-(E+E')^2}.
\label{Higgs_q_sub}
\end{eqnarray}
Here, we have replaced $i\Omega_n\to\omega$.
At $\bm q=0$, Eq.~(\ref{Higgs_q}) reduces to
\begin{eqnarray}
&&\frac{y}{N}\sum_{\bm p}\frac{1}{E}\frac{1}{\omega^2-4E^2}\nonumber\\
&&=y\int_0^{3t}d\varepsilon\
 \frac{\rho(\varepsilon)}{\sqrt{\varepsilon^2+\Delta^2}[\omega^2-4(\varepsilon^2+\Delta^2)]}=0,
\label{Higgs_q0}
\end{eqnarray}
where $y=\omega^2/4-\Delta^2$ and 
$\rho(\varepsilon)=\frac{1}{N}\sum_{\bm p}\delta(\varepsilon-|\gamma_{\bm p}|)$
is the density of states of the fermion energy band. If we set
$\omega=2\Delta$, the denominator in the integrand of
Eq.~(\ref{Higgs_q0}) is proportional to $\varepsilon^2$, while the
numerator is proportional to $\varepsilon$ for small $\varepsilon$
because $\rho(\varepsilon)\propto \varepsilon$.
The integral in Eq.~(\ref{Higgs_q0})
is thus well defined in the limit $\omega\to 2\Delta$.
In this limit, Eq.~(\ref{Higgs_q0}) is satisfied when $y=0$ and consequently the energy of the
Higgs mode is obtained as $\omega_{\rm Higgs}(\bm q=0)=2\Delta$.
\par
To derive the energy dispersion for small $\bm q$, we expand
Eq.~(\ref{Higgs_q}) to second order in ${\bm q}$. 
Using the relations
\begin{widetext}
\begin{eqnarray}
&&\gamma_{\bm p+\bm q}\simeq\gamma_{\bm
 p}+\delta\gamma_1+\delta\gamma_2,\ 
\delta\gamma_1=-it\{e^{i\bm p\cdot\bm a_1}(\bm q\cdot\bm a_1)+e^{i\bm
p\cdot\bm a_2}(\bm q\cdot\bm a_2)\},\\  
&&\delta\gamma_2=\frac{t}{2}\{e^{i\bm p\cdot\bm a_1}(\bm q\cdot\bm a_1)^2+e^{i\bm
p\cdot\bm a_2}(\bm q\cdot\bm a_2)^2\},\\
&&w_1={\rm Re}[\gamma^*\delta\gamma_1]=-t^2\left\{(\sin(\bm p\cdot\bm a_1)+\sin(\bm p\cdot\bm a_3))(\bm q\cdot\bm a_1)+(\sin(\bm p\cdot\bm a_2)-\sin(\bm p\cdot\bm a_3))(\bm q\cdot\bm a_2)\right\},\\
&&w_1'={\rm Im}[\gamma^*\delta\gamma_1]=t^2\left\{(1+\cos(\bm p\cdot\bm a_1)+\cos(\bm p\cdot\bm
	    a_3))(\bm q\cdot\bm a_1)+(1+\cos(\bm p\cdot\bm a_2)+\cos(\bm p\cdot\bm
	    a_3))(\bm q\cdot\bm a_2)\right\},\\
&&w_2={\rm Re}[\gamma^*\delta\gamma_2]=-\frac{t^2}{2}\left\{(1+\cos(\bm p\cdot\bm a_1)+\cos(\bm p\cdot\bm
	    a_3))(\bm q\cdot\bm a_1)^2+(1+\cos(\bm p\cdot\bm a_2)+\cos(\bm p\cdot\bm
	    a_3))(\bm q\cdot\bm a_2)^2\right\},\\
&&w_2'={\rm Im}[\gamma^*\delta\gamma_2]=-\frac{t^2}{2}\left\{(\sin(\bm p\cdot\bm a_1)+\sin(\bm p\cdot\bm
	    a_3))(\bm q\cdot\bm a_1)^2+(\sin(\bm p\cdot\bm a_2)-\sin(\bm p\cdot\bm
	    a_3))(\bm q\cdot\bm a_2)^2\right\},\\
&&|\gamma_{\bm p+\bm q}|^2\simeq |\gamma_{\bm p}|^2+s_1+s_2,\ \
 s_1=2w_1,\ s_2=|\delta\gamma_1|^2+2w_2,\\
&&|\delta\gamma_1|^2=t^2\left\{(\bm q\cdot\bm a_1)^2+2\cos(\bm p\cdot\bm a_3)(\bm q\cdot\bm a_1)(\bm
	q\cdot\bm a_2)+(\bm q\cdot\bm a_2)^2\right\},
\end{eqnarray}
\end{widetext}
one finds that Eq.~(\ref{Higgs_q_sub}) becomes
\begin{eqnarray}
&&X\simeq\frac{1}{2N}\sum_{\bm
 p}\frac{2}{E}\frac{\omega^2-(2|\gamma|^2+s_1+s_2)-4\Delta^2}{\omega^2-4E^2}\nonumber\\
&&=\frac{1}{N}\sum_{\bm
 p}\frac{1}{E}\frac{\omega^2-(2|\gamma|^2+s_2)-4\Delta^2}{\omega^2-4E^2}.
\label{X}
\end{eqnarray}
Since the factor $\sin(\bm p\cdot\bm a_i)$ is odd for $\bm p$, the summation for
$\bm p$ including this factor vanishes. Similarly, $R$ can be approximated as 
\begin{eqnarray}
&&R\simeq-\frac{1}{N}\sum_{\bm
 p}\frac{2}{E}\frac{(\gamma+\delta\gamma_1+\delta\gamma_2)\gamma^*}{\omega^2-4E^2},\\
&&{\rm Re}R=-\frac{1}{N}\sum_{\bm
 p}\frac{2}{E}\frac{|\gamma|^2+w_2}{\omega^2-4E^2}=R_0+R_2,\label{Ref}\\
&&{\rm Im}R=-\frac{1}{N}\sum_{\bm p}\frac{2}{E}\frac{w_1'}{\omega^2-4E^2}=R_1.\label{Imf}
\end{eqnarray}
\par
In evaluating further Eqs.~(\ref{X}), (\ref{Ref}), and (\ref{Imf}), we
encounter the factor
\begin{equation}
\sum_{\bm p}\frac{1}{E}\frac{\cos(\bm p\cdot\bm a_i)}{\omega^2-4E^2}.
\end{equation}
Here, we replace $\cos(\bm p\cdot\bm a_i)$ in the integrand with its value
at the $K$ ($K'$) point, i.e., $\langle \cos(\bm p\cdot\bm a_i)\rangle=\cos(\bm
p_{K}\cdot\bm a_i)=-1/2$. At half-filling, since the Fermi level is at the $K$
($K'$) point, this replacement is justified for small $q$. 
As a result, $R_1$ and $R_2$ vanish and we finally obtain
\begin{equation}
X-|R|\simeq \left(4y-v_F^2q^2\right) \frac{1}{N}\sum_{\bm
 p}\frac{1}{E}\frac{1}{\omega^2-4E^2}=0,\label{X-R}
\end{equation}
where $v_F=3t/2$ is the Fermi velocity.
Consequently, the dispersion of the Higgs mode is obtained as
\begin{equation}
\omega_{\rm Higgs}^2=4\Delta^2+v_F^2 q^2.
\end{equation}
In the limit $\Delta^2\gg v_F^2 q^2$, $\omega_{\rm Higgs}$ is approximated as
$\omega_{\rm Higgs}= 2\Delta+v_F^2q^2/4\Delta$ which is plotted in Fig.~1 (f).
On the other hand, at the transition point with $\Delta=0$, the
dispersion for small $q$ coincides with that of the lower edge of the continuum as
$\omega_{\rm Higgs}= v_Fq.$

\subsection{Energy dispersion of the AB/Leggett mode}

We derive the energy dispersion of the AB/Leggett mode for small momentum.
Subtracting Eq.~(\ref{gapeq2}) from $1+U[(C-D)-\sqrt{4F^2+|R|^2}]=0$
which is equivalent to Eq.~(11), we obtain
\begin{equation}
U\left(Y-\sqrt{4F^2+|R|^2}\right)=0,
\label{AB_q}
\end{equation}
where 
\begin{eqnarray}
&&Y=(C-D)+\frac{1}{2N}\sum_{\bm p}\frac{E+E'}{EE'}\nonumber\\
&&=\frac{1}{2N}\sum_{\bm
 p}\frac{E+E'}{EE'}\frac{\omega^2-(|\gamma|^2+|\gamma'|^2)}{\omega^2-(E+E')^2}.
\label{AB_q_sub}
\end{eqnarray}
Setting $\bm q=0$, we obtain
\begin{eqnarray}
Y&=&\frac{1}{N}\sum_{\bm
 p}\frac{1}{E}\frac{\omega^2-2|\gamma|^2}{\omega^2-4E^2}=d_1\omega^2-2d_2,\\
R&=&-\frac{2}{N}\sum_{\bm
 p}\frac{1}{E}\frac{|\gamma|^2}{\omega^2-4E^2}=-2d_2,\\
F&=&\frac{1}{N}\sum_{\bm p}\frac{1}{E}\frac{\Delta\omega}{\omega^2-4E^2}=d_2\Delta\omega,
\end{eqnarray}
where
\begin{eqnarray}
d_1(\omega)=\frac{1}{N}\sum_{\bm p}\frac{1}{E}\frac{1}{\omega^2-4E^2},\\
 \ d_2(\omega)=\frac{1}{N}\sum_{\bm p}\frac{1}{E}\frac{|\gamma|^2}{\omega^2-E^2}.
\end{eqnarray}
Thus, Eq.~(\ref{AB_q}) reduces to
\begin{eqnarray}
&&Y-\sqrt{4F^2+|R|^2}\nonumber\\
&&=\frac{d_1\omega^2(d_1\omega^2-4d_2-4d_1\Delta^2)}{(d_1\omega^2-2d_2)+\sqrt{4d_1^2\Delta^2\omega^2+4d_2^2}}=0.
\label{Yq0}
\end{eqnarray}
From $d_{i=1,2}(\omega<2\Delta)<0$, we obtain the gapless AB/Leggett mode: $\omega_{\rm AB}(\bm q=0)=0$. 
Note that the term within parentheses in the numerator of Eq.~(\ref{Yq0}) at $\omega=0$ is found to give
\begin{eqnarray}
-4d_2(\omega=0)-4d_1(\omega=0)\Delta^2=\frac{1}{N}\sum_{\bm
 p}\frac{1}{E}=\frac{1}{U}.
\label{d_2}
\end{eqnarray}
\par
For small $\bm q$, expanding $F$, $R$, and $Y$ to second order in
$\bm q$, we obtain
\begin{eqnarray}
&&Y\simeq d_1\omega^2-2d_2-d_1 v_F^2 q^2,\label{Yq}\\
&&F\simeq \Delta d_1\omega,\ \
R\simeq -2d_2.\label{FRq}
\end{eqnarray}
In Eqs.~(\ref{Yq}) and (\ref{FRq}), we used the same approximation as the one for Eq.~(\ref{X-R}). As a result, we obtain
\begin{eqnarray}
&&Y-\sqrt{4F^2+|R|^2}\nonumber\\
&&\simeq\frac{d_1(\omega=0)(\omega^2/U+4d_2(\omega=0) v_F^2
 q^2)}{(d_1\omega^2-2d_2-2d_1
 v_F^2q^2)+\sqrt{4\Delta^2d_1^2\omega^2+4d_2^2}}\nonumber\\
&&=0. 
\end{eqnarray}
We have used Eq.~(\ref{d_2}) to derive the final expression. The pole is thus given by
\begin{eqnarray}
&&\omega_{\rm AB}=\lambda v_F q,\\ 
&&\lambda^2=4U|d_2(\omega=0)|=\frac{U}{N}\sum_{\bm p}\frac{|\gamma|^2}{E^3}.
\end{eqnarray}
Note that $\lambda\leq 1$ from the gap equation (2). 
At the transition point
($\Delta=0$), $\lambda=1$ and thus the AB/Leggett mode becomes degenerate with the Higgs mode as well as
the edge of the continuum as $\omega_{\rm AB}=v_F q$.

\end{document}